\documentclass[prd,nofootinbib,notitlepage,10pt,aps]{revtex4-1}
\usepackage{epsfig,amsmath,amssymb,slashed}
\usepackage[utf8]{inputenc}
\usepackage{hyperref}
\usepackage{natbib}
\usepackage{amsthm}
\usepackage{xcolor}
\usepackage[normalem]{ulem}

\newcommand{\mean}[1]{\langle #1 \rangle}

\newcommand{\di}{{\rm d}}
\newcommand{\D}{{\rm d}}
\newcommand{\ii}{i}
\newcommand{\I}{i}

\def\wT{{\widehat T}}
\def\wj{{\widehat j}}
\def\wQ{{\widehat Q}}
\def\wO{{\widehat O}}
\def\wP{{\widehat P}}
\def\wJ{{\widehat J}}

\def\wspt{{\widehat{\cal S}}}

\def\wrho{{\widehat{\rho}}}

\def\wrhol{{\widehat{\rho}_{\rm LE}}}

\newcommand{\tr}{{\rm tr}}  
\newcommand{\Tr}{{\rm Tr}}

\newcommand{\e}{{\rm e}}

\newcommand{\p}{{\rm p}}

\newcommand{\Psibar}{{\overline \Psi}}
\newcommand{\E}{{\rm e}}
\newcommand{\de}{\partial}
\newcommand{\ped}[1]{_{\textup{#1}}}

\newcommand{\be}{\begin{equation}}
\newcommand{\ee}{\end{equation}}                                                                               
\newcommand{\bea}{\begin{eqnarray}}
\newcommand{\eea}{\end{eqnarray}}

\newcommand{\h}[1]{\widehat{#1}}
\renewcommand{\vec}[1]{\ensuremath{\mathchoice				
                     {\mbox{\boldmath$\displaystyle\mathbf{#1}$}}
                     {\mbox{\boldmath$\textstyle\mathbf{#1}$}}
                     {\mbox{\boldmath$\scriptstyle\mathbf{#1}$}}
                     {\mbox{\boldmath$\scriptscriptstyle\mathbf{#1}$}}}}

\begin{document}

\title{Spin-thermal shear coupling in a relativistic fluid}

\author{F. Becattini}\email{becattini@fi.infn.it}\affiliation{Universit\`a di 
 Firenze and INFN Sezione di Firenze, Via G. Sansone 1, 
	I-50019 Sesto Fiorentino (Florence), Italy} 
\author{M. Buzzegoli}\email{matteo.buzzegoli@unifi.it}\affiliation{Universit\`a di 
 Firenze and INFN Sezione di Firenze, Via G. Sansone 1, 
	I-50019 Sesto Fiorentino (Florence), Italy}
 \author{A. Palermo}\email{andrea.palermo@unifi.it}\affiliation{Universit\`a di 
 Firenze and INFN Sezione di Firenze, Via G. Sansone 1, 
	I-50019 Sesto Fiorentino (Florence), Italy}

\begin{abstract}
We show that spin polarization of a fermion in a relativistic fluid at local thermodynamic 
equilibrium can be generated by the symmetric derivative of the four-temperature vector, defined
as thermal shear. As a consequence, besides vorticity, acceleration and temperature gradient,
also the shear tensor contributes to the polarization of particles in a fluid. This contribution 
to the spin polarization vector, which is entirely non-dissipative, adds to the well known term 
proportional to thermal vorticity and may thus have important consequences for the solution of 
the local polarization puzzles observed in relativistic heavy ion collisions.
\end{abstract}

\maketitle
\section{Introduction}
\label{intro}

In a rotating fluid at global thermodynamic equilibrium, particle spin gets polarized along
the direction of the angular velocity vector by an amount which is proportional to $\hbar \omega/KT$.
This phenomenon is the essence of the Barnett effect \cite{barnett} and it has been known for a long 
time. In a relativistic fluid at local thermodynamic equilibrium, the covariant form of statistical
mechanics dictates that spin polarization is driven by {\em thermal vorticity}:
\be\label{thvort}
  \varpi_{\mu\nu} = -\frac{1}{2} \left( \partial_\mu \beta_\nu - \partial_\nu \beta_\mu \right)
\ee
where $\beta$ is the four-temperature vector:
\be\label{fourtemp}
  \beta^\mu = \frac{1}{T} u^\mu ,
\ee  
$u$ being the four-velocity and $T$ the proper temperature. At first order in thermal vorticity, the 
formula relating the mean spin vector $S^\mu(p)$ of a spin $1/2$ fermion to thermal vorticity reads 
\cite{Becattini:2013fla}:
\be\label{basic}
 S^\mu(p)= - \frac{1}{8m} \epsilon^{\mu\rho\sigma\tau} p_\tau 
 \frac{\int_{\Sigma} \di \Sigma \cdot p \; n_F (1 -n_F) \varpi_{\rho\sigma}}
  {\int_{\Sigma} \di \Sigma \cdot p \; n_F}
\ee
where $\Sigma$ is a 3D hypersurface, $n_F$ is the Fermi-Dirac phase-space distribution function:
$$
  n_F = \frac{1}{\exp[\beta\cdot p - q \mu/T]+1},
$$
$q$ being the charge of the particle and $\mu$ the corresponding chemical potential. The equation
\eqref{basic} predicts that a particle can get a spin polarization in the presence of gradients of 
temperature, vorticity and acceleration. 

The observation of spin polarization in relativistic nuclear collisions \cite{star} confirmed
the predictions of the formula~\eqref{basic} for the global polarization (with $\Sigma$ the hadronization
hypersurface), that is integrated over all momenta. In fact, the formula \eqref{basic} failed to reproduce 
the measurements as a function of momentum \cite{Adam:2019srw,Niida:2018hfw,Becattini:2020ngo}; particularly, 
the sign of the longitudinal polarization and the polarization along
the angular momentum as a function of the azimuthal angle, which has been investigated in several
papers \cite{Becattini:2015ska,Xia:2018tes,Florkowski:2019qdp,Florkowski:2019voj,Liu:2019krs,Fu:2020oxj,Xie:2019jun,Sun:2018bjl,Wu:2019eyi}.

In this work, we will show that also the symmetric gradient of $\beta$ contributes to the 
spin at local thermodynamic equilibrium at the leading order. This term is non-dissipative as 
well as non-local for it depends on a specific 3D integration hypersurface and implies a new 
relativistic effect, namely a coupling between spin and the shear tensor in a relativistic 
fluid.

\subsection*{Notation}
In this paper we adopt the natural units, with $\hbar=c=K=1$. The Minkowskian metric tensor $g$ is ${\rm diag}(1,-1,-1,-1)$; for the Levi-Civita symbol we use the convention $\epsilon^{0123}=1$.\\ 
We will use the relativistic notation with repeated indices assumed to be saturated. Operators in Hilbert 
space will be denoted by a wide upper hat, e.g. $\widehat H$, except the Dirac field operator which is
denoted by a $\Psi$.

\section{Local thermodynamic equilibrium and its gradient expansion}
\label{sec:local}

\begin{figure}[t!hb]
\centering
\includegraphics{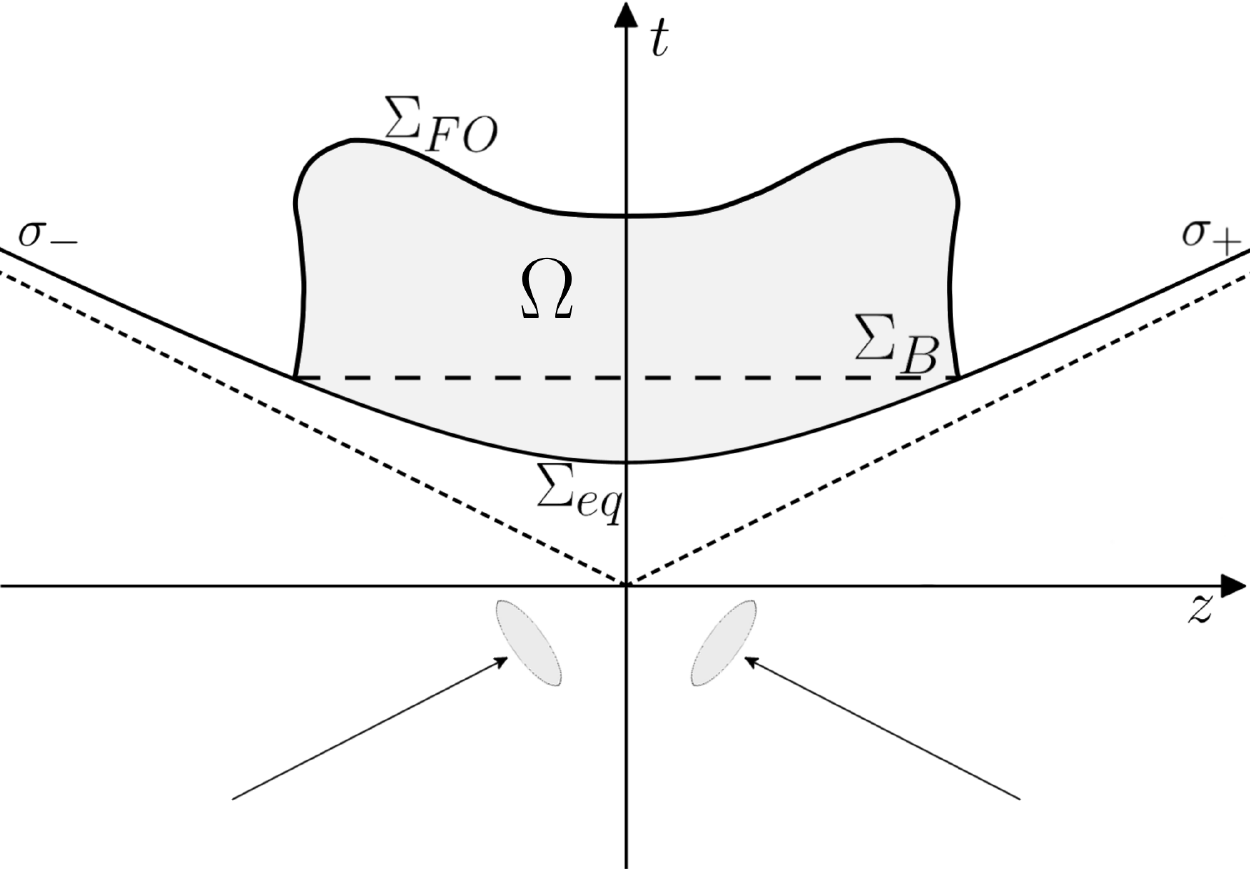}
\caption{The space-time diagram of a relativistic nuclear collision in the
center-of-mass frame. $\Sigma_{eq}$ is the 3D hypersurface where LTE is achieved, $\Sigma\ped{FO}$ is the freeze-out
hypersurface. The $\sigma_\pm$ are the side branches subsets of
$\Sigma_{eq}$ and $\Sigma\ped{B}$ is the hyperplane connecting the
limiting surfaces of $\Sigma_{FO}$. In the volume $\Omega$, matter is in the quark-gluon plasma phase.}
\label{figure}
\end{figure}

For a relativistic quantum fluid which, at some time, achieves Local Thermodynamic Equilibrium (LTE), 
a powerful approach is the Zubarev's method of the stationary non-equilibrium density operator 
\cite{Zubarev:1979,vanWeert}. We refer the reader to the recent paper \cite{Becattini:2019dxo} for a more 
detailed description. The actual density operator of such a fluid, in the Heisenberg representation, is:
\be\label{densop}
  \wrho = \dfrac{1}{Z} 
  \exp \left[ -\int_{\Sigma_{eq}} \di \Sigma_\mu \left( \wT^{\mu\nu}(x) 
  \beta_\nu(x) - \zeta(x) \wj^\mu(x) \right) \right],
\ee
where $\beta$ is the four-temperature vector, $\zeta$ the ratio between chemical potential and 
temperature and $\Sigma_{eq}$ is some initial 3D hypersurface where LTE is achieved. For relativistic 
nuclear collisions, this is supposedly the 3D hyperbolic hypersurface where the quark-gluon plasma (QGP) 
thermalizes (see figure~\ref{figure}). It should be pointed out that the form of the local equilibrium density 
operator is pseudo-gauge dependent \cite{Becattini:2018duy,Speranza:2020ilk}, with the form 
in eq.~\eqref{densop} applying to the Belinfante stress-energy tensor only. We note right away that 
the final result of this work would be the same if we used the canonical stress-energy tensor instead;
this is shown in detail in the Appendix \ref{sec:canonical}. Henceforth, it will be understood that 
$\wT$ is the Belinfante symmetrized stress-energy tensor.

The operator \eqref{densop} can be turned into a more manageable form by means of the Gauss' theorem, 
taking into account that $\wT$ and $\wj$ are conserved currents:
\be\label{nedo}
 \wrho =  \dfrac{1}{Z} 
 \exp\left[ - \int_{\Sigma_{eq}} \!\!\!\!\!\! \di \Sigma_\mu \; \left( \wT^{\mu\nu} 
  \beta_\nu - \wj^\mu \zeta \right) \right] =
 \dfrac{1}{Z} 
 \exp\left[ - \int_{\Sigma(\tau)} \!\!\!\!\!\! \di \Sigma_\mu \; \left( \wT^{\mu\nu} 
  \beta_\nu - \wj^\mu \zeta \right) + \int_\Omega \di \Omega \; \left( \wT^{\mu\nu} 
  \nabla_\mu \beta_\nu - \wj^\mu \nabla_\mu \zeta \right) \right] ,
\ee
where $\Sigma(\tau)$ is some 3D hypersurface at ``present'' time $\tau$. In the case of heavy ion collisions 
the hypersurface $\Sigma(\tau)$ is usually the joining of the freeze-out hypersurface $\Sigma_{FO}$
encompassing the QGP space-time region and the two side branches $\sigma_\pm$ subsets of the 
$\Sigma_{eq}$, as shown in fig. \ref{figure}. A peculiarity of the heavy ion collisions is that 
the hypersurface of ``present'' local equilibrium is partly time-like, that is $\hat n \cdot \hat n = -1$.
 
In the right hand side of the density operator \eqref{nedo} the first term is the LTE and, 
as expected for a quasi-ideal fluid such as the QGP, it is the predominant one; the second term is, on 
the other hand, supposedly a correction and it is responsible for everything that can be called 
dissipative. Indeed, it can 
be shown that entropy is generated only if the second term is non-vanishing \cite{Zubarev:1979}.

We shall focus on the LTE term:
\be\label{rhole}
  \wrhol = \frac{1}{Z_{\rm LE}} \exp \left[ - \int_\Sigma \di \Sigma_\mu \; 
  \left( \wT^{\mu\nu}(y) \beta_\nu(y) - \wj^\mu(y) \zeta(y) \right) \right] .
\ee
The density operator~\eqref{rhole}, as well as eqs.~\eqref{densop} and~\eqref{nedo}, is independent 
of the hypersurface only if the $\beta$ field satisfies the Killing equation \cite{Becattini:2012tc}.
When calculating the mean values of any local operator $\widehat O(x)$, as the thermodynamic fields 
$\beta$ and $\zeta$ are supposedly slowly varying compared to the correlation lengths between $\wO$ 
and the operators $\wT$ and $\wj$ (see e.g. ref.~\cite{Becattini:2015nva}), it is a good approximation 
to expand $\beta$ and $\zeta$ in a Taylor series from $x$:
$$
 \beta_\nu(y) \simeq \beta_\nu(x) + \partial_\lambda \beta_\nu(x) (y-x)^\lambda
$$
and similarly for $\zeta$. Thus, we have:
\begin{align}\label{expvalue}
\Tr&(\wrhol \wO(x))  \\ \nonumber
 \simeq& \frac{1}{Z_{\rm LE}}
  \Tr \left( \exp \left[ - \int_\Sigma \di \Sigma_\mu \; 
  \left( \wT^{\mu\nu}(y) [ \beta_\nu(x) + \partial_\lambda\beta_\nu(x)(y-x)^\lambda ] 
  - \wj^\mu(y) [\zeta(x) +\partial_\lambda \zeta(x) (y-x)^\lambda] \right) \right] 
  \wO(x) \right) \\ \nonumber
 =& \frac{1}{Z_{\rm LE}} \Tr \left( \exp \left[ - \beta_\nu(x) \int_\Sigma 
  \di \Sigma_\mu \; \wT^{\mu\nu}(y) - \partial_\lambda \beta_\nu(x) 
 \int_\Sigma \di \Sigma_\mu \; (y-x)^\lambda \wT^{\mu\nu}(y) 
  - \zeta(x) \int_\Sigma  \di \Sigma_\mu \;\wj^\mu +\right.\right. \\ \nonumber
 & \left.\left.  - \partial_\lambda \zeta(x) \int_\Sigma \di \Sigma_\mu \; 
  (y-x)^\lambda \wj^\mu \right] \wO (x) \right) .
\end{align}
For the sake of simplicity, we will omit the gradients of $\zeta$ and focus on the gradients
of $\beta$, which are the most relevant for our purposes. These gradients can be split into a 
symmetric and an anti-symmetric part giving rise to:
\be\label{symmanti}
  \frac{1}{2} \varpi_{\lambda\nu} \int_\Sigma \di \Sigma_\mu \; 
  \left[(y-x)^\lambda \wT^{\mu\nu}(y) - (y-x)^\nu \wT^{\mu\lambda}(y)\right]
 - \frac{1}{4} (\partial_\lambda \beta_\nu + \partial_\nu \beta_\lambda) 
  \int_\Sigma \di \Sigma_\mu \; 
    \left[(y-x)^\lambda \wT^{\mu\nu}(y) + (y-x)^\nu \wT^{\mu\lambda}(y)\right]
\ee
where $\varpi$ is the thermal vorticity \eqref{thvort}. We can recognize in the first term 
of the above equation the total angular momentum-boost operator $\wJ^{\lambda\nu}_x$ 
(with a proviso, see ref.~\cite{Becattini:2020sww})
centered in $x$, while the second term includes the non-conserved operator:
\be\label{quadr}
  \wQ^{\lambda\nu}_x = \int_\Sigma \di \Sigma_\mu \; \left[(y-x)^\lambda \wT^{\mu\nu}(y) + 
  (y-x)^\nu \wT^{\mu\lambda}(y)\right]
\ee
coupled to the {\em thermal shear tensor}:
\be\label{thshear}
  \xi_{\lambda\nu} = \frac{1}{2} (\partial_\lambda \beta_\nu + \partial_\nu \beta_\lambda) 
\ee
which vanishes at global thermodynamic equilibrium due to the Killing condition. 
It is very important to stress that $\wQ_x$ is a tensor in a more limited sense than the 
angular momentum-boost operator $\wJ_x$. Indeed, since the integrand of $\wQ_x$ in eq.~\eqref{quadr}
is not divergenceless:
$$
 \partial_\mu \left[(y-x)^\lambda \wT^{\mu\nu}(y) + (y-x)^\nu \wT^{\mu\lambda}(y)\right]
  = 2 \wT^{\lambda\nu}
$$
its value specifically depends on the hypersurface $\Sigma$, unlike $\wJ_x$, and, strictly speaking, 
should then be denoted as $\wQ_x(\Sigma)$ (even though we will not use that notation). In quantum 
language, the operator $\wQ_x$ does not fulfill the transformation rule for a tensor operator under a 
Lorentz transformation, that is:
$$
  \widehat\Lambda \, \wQ_x^{\mu\nu} {\widehat\Lambda}^{-1} \ne \Lambda^{-1 \mu}_{\quad \rho} 
   \Lambda^{-1 \nu}_{\quad \sigma} \wQ_x^{\rho\sigma} \, ,
$$
where $\widehat \Lambda$ is the unitary representation of the Lorentz transformation $\Lambda$ in 
the Hilbert space. This implies that all results involving $\wQ_x$, for instance in quantum 
correlators, will eventually depend on that hypersurface and are thus expected to break local covariance.

Altogether, in the equation~\eqref{expvalue}, we can approximate the local 
thermodynamic equilibrium operator at first order in the gradients as:
\be\label{rholeappr}
 \wrhol \simeq \frac{1}{Z_{\rm LE}} \exp \left[ - \beta_\nu(x) 
 \wP^\nu + \frac{1}{2} \varpi_{\lambda\nu}(x) \wJ^{\lambda\nu}_x - 
 \frac{1}{2} \xi_{\lambda\nu}(x) \wQ^{\lambda\nu}_x(\Sigma) \right]. 
\ee
where $\wP$ is the total four-momentum operator and the dependence on the hypersurface 
of the last term was highlighted.

\section{Spin and thermal shear tensor}

The mean spin polarization vector of a spin $1/2$ particle can be obtained from the particle term (i.e. 
the future time-like part) of the Wigner function $W^+$~\cite{Becattini:2020sww}:
\begin{equation}\label{eq:PolVec}
S^\mu(k)=\frac{1}{2}\frac{\int_\Sigma \D\Sigma \cdot k\,\tr\left[\gamma^\mu\gamma^5 W^+(x,k)\right]}
    {\int_\Sigma \D\Sigma\cdot  k\, \tr\left[W^+(x,k)\right]}.
\end{equation}
As pointed out in ref.~\cite{Becattini:2020sww}, this formula applies to free, or 
quasi-free fields, therefore, in relativistic nuclear collisions, only to hadronic fields if $\Sigma$ 
is a 3D hypersurface outside $\Sigma_{FO}$ in fig.~\ref{figure}. Furthermore, it is convenient to
set $\Sigma=\Sigma_{FO}$ to calculate it most accurately. The Wigner function is the expectation 
value of the Wigner operator \cite{DeGroot:1980dk}:
\begin{equation}\label{eq:wigop}
\h{W}^+_{ab}(x,k) =  \theta(k^0)\theta(k^2)\frac{1}{(2\pi)^4} \int \D^4 s \; \E^{-\I k \cdot s}
   :\Psibar_b (x+s/2) \Psi_a (x-s/2): \, ,
\end{equation}
$\Psi$ being the free Dirac field, ``$:\,:$'' denotes the normal ordering and $\theta$ the Heaviside
step function. In the application to relativistic nuclear collisions, $\Psi$
is to be understood as an effective hadronic field. For our purposes, the Wigner function is the 
expectation value of \eqref{eq:wigop} with the LTE operator \eqref{rhole} 
$W^+_{ab}(x,k)=\Tr\left(\wrhol \h{W}^+_{ab}(x,k) \right)$.
As has been mentioned, in the hydrodynamic limit of slowly varying thermodynamic fields, one can 
approximate the Wigner function in $x$ by making a Taylor expansion of the four-temperature 
in~\eqref{rhole} around the point $x$. The contribution to the local expectation values of the gradient 
terms of the \eqref{rholeappr} is small compared to the contribution from the term $\beta(x) \cdot \wP$, 
hence one can expand the exponential in \eqref{rholeappr} with the familiar techniques of linear response theory~\cite{vanWeert}:
\begin{equation*}
\e^{\h{A}+\h{B}}=\e^{\h{A}}+\int_0^1 \di z\,\e^{z \h{A}}\,\h{B}\,\e^{-z \h{A}}
\,\e^{\h{A}}+\cdots,
\end{equation*}
where:
\begin{equation*}
\h{A}=-\beta(x)\cdot \h{P},\quad
\h{B} = \frac{1}{2} \varpi_{\nu\lambda}(x) \wJ^{\nu\lambda}_x - \frac{1}{2} \xi_{\mu\nu}(x) 
 \wQ^{\mu\nu}_x \, .
\end{equation*}
Thereby, the Wigner function, and the spin vector as well, in \eqref{eq:PolVec} will receive two linear 
corrections: one proportional to thermal vorticity involving correlators between the Wigner operator 
and the angular momentum-boost operator $\wJ_x$ and one proportional to the thermal shear tensor, involving 
correlators between the Wigner operator and the operator $\wQ_x$.
The thermal shear tensor contribution to the spin vector has been usually neglected, for a twofold reason: 
first, it certainly vanishes at global equilibrium because of Killing condition and, secondly, the 
symmetric part of the gradients are cancelled by the Levi-Civita tensor in the formula \eqref{basic}. 
However, we will show, that a combination of thermal shear tensor and momenta eventually survives 
and gives rise to a contribution which can be numerically important, especially for a fluid which is not 
yet very close to global equilibrium.

To make calculations more compact, we will not separate the symmetric and antisymmetric part of the Taylor
expansion of $\beta$ like in eq.~\eqref{symmanti} and study the linear response in terms of the 
perturbation:
$$
\h{B}=-\int_{\Sigma}\D\Sigma_\lambda(y)\h{T}^{\lambda\rho}(y)\Delta\beta_\rho(x,y) ,
$$
where:
\be\label{deltabeta}
\Delta\beta_\rho(x,y)= \beta_\rho(y) -\beta_\rho(x) \simeq \partial_\sigma \beta_\rho(x) (y-x)^\sigma .
\ee
Therefore, the particle term of the Wigner function at LTE can be approximated by:
\begin{equation}\label{eq:WigFunLE}
\mean{\h{W}^+_{ab}(x,k)}\ped{LE}\simeq \mean{\h{W}^+_{ab}(x,k)}_{\beta(x)}
    + \Delta W^+_{ab}(x,k),
\end{equation}
with:
\begin{equation}\label{eq:WigfunCorr}
\Delta W^+_{ab}(x,k) = -\int_0^1\D z\int_{\Sigma}\D\Sigma_\lambda(y)\Delta\beta_\rho(x,y)
    \mean{\h{W}^+_{ab}(x,k)\h{T}^{\lambda\rho}(y+\I z\beta(x))}_{c,\beta(x)}
\end{equation}
where with $\mean{\cdots}_{\beta(x)}$ we denote the thermal expectation values calculated at the
homogeneous global thermodynamic equilibrium, i.e. with the density operator:
$$
\h{\rho}_0 = \frac{1}{Z} \exp[-\beta(x) \cdot \h{P}].
$$
The subscript $c$ on the thermal average in~\eqref{eq:WigfunCorr} denotes the connected part of 
the correlator, that is, for the simplest case of two operators:
$$
\mean{\h{O}_1 \h{O}_2}_c\equiv \mean{\h{O}_1 \h{O}_2}-\mean{\h{O}_1}\mean{ \h{O}_2}.
$$

The two terms of the right hand side of the eq.~\eqref{eq:WigFunLE} can be evaluated with
standard techniques (see Appendix~\ref{sec:AppWigFun}) and turn out to be:
\begin{equation*}\label{eq:wigner0}
W^+_0(x,k)=\mean{\h{W}^+(x,k)}_{\beta(x)}=\left(m+\gamma^\mu k_\mu\right)
    \delta(k^2-m^2)\theta(k_0)\frac{1}{(2\pi)^3} n\ped{F}\left( k\right)
\end{equation*}
and
\begin{equation}\label{eq:DeltaWigPriv}
\begin{split}
\Delta W_{ab}^+(x,k)=&-\int_0^1\D z\int_{\Sigma}\D\Sigma_\lambda(y)\Delta\beta_\rho(x,y)
\frac{1}{(2\pi)^6}\int\frac{\D^3\p}{2\varepsilon_p}\int\frac{\D^3\p'}{2\varepsilon_{p'}}
\delta^4\left(k-\frac{p+p'}{2} \right)\times\\
&\mathcal{T}^{\lambda\rho}(p,p')_{ab}\,
\E^{\I(p-p')(x-y)}\E^{z(p-p')\beta}n\ped{F}(p)(1-n\ped{F}(p')),
\end{split}
\end{equation}
where $\mathcal{T}$ is
\be\label{tb}
\mathcal{T}^{\lambda\rho}(p,p')_{ab}=
\frac{1}{4}\left[(\slashed{p}'+m)\gamma^\lambda(\slashed{p}+m) \right]_{ab}
(p^\rho+p^{\prime\rho})+\frac{1}{4}\left[(\slashed{p}'+m)\gamma^\rho(\slashed{p}+m) \right]_{ab}
(p^\lambda+p^{\prime\lambda}).
\ee
We can now replace the $\Delta \beta$ in eq.~\eqref{eq:DeltaWigPriv} by using eq.~\eqref{deltabeta}:
\begin{equation} \label{eq:DeltaWig}
\begin{split}
\Delta W_{ab}^+(x,k)=&-\int_0^1\D z\int_{\Sigma}\D\Sigma_\lambda(y)\de_\kappa\beta_\rho
\frac{(y-x)^\kappa}{(2\pi)^6}\int\frac{\D^3\p}{2\varepsilon_p}\int\frac{\D^3\p'}{2\varepsilon_{p'}}
\delta^4\left(k-\frac{p+p'}{2} \right)\\
&\times\mathcal{T}^{\lambda\rho}(p,p')_{ab}\,
\E^{\I(p-p')(x-y)}\E^{z(p-p')\beta}n\ped{F}(p)(1-n\ped{F}(p')).
\end{split}
\end{equation}
It is convenient  to work out the integration over $\Sigma$ in the eq.~\eqref{eq:DeltaWig} by using the Gauss' 
theorem, and splitting it into an integral over a flat 3D hypersurface $\Sigma_B$ and a 4D integral over 
the region $\Omega_B$ encompassed by $\Sigma$ and $\Sigma_B$; for instance, for heavy ion collisions 
applications, it is convenient to choose $\Sigma=\Sigma_{FO}$, see fig.~\ref{figure}. The formula 
\eqref{eq:PolVec} is thus the sum of a 4D integral and a 3D boundary term:
\begin{equation}\label{eq:PolVandB}
S^\mu(k)\simeq  S_{\de\beta,\Omega}^\mu(k) +S_{\de\beta,B}^\mu(k) =
\frac{1}{2}\frac{\int_\Sigma \di\Sigma \cdot k\,\tr\left[\gamma^\mu\gamma^5
\Delta_\Omega W^+(x,k)\right]}{\int_\Sigma \D\Sigma \cdot  k\, \tr\left[W^+_0(x,k)\right]}
+\frac{1}{2}\frac{\int_\Sigma \di\Sigma \cdot k\,\tr\left[\gamma^\mu\gamma^5
\Delta_B W^+(x,k)\right]}{\int_\Sigma \D\Sigma \cdot  k\, \tr\left[W^+_0(x,k)\right]},
\end{equation}
where:
\begin{equation*}
\begin{split}
\Delta_\Omega W_{ab}^+(x,k)=&-\int_0^1\D z\int_{\Omega_B}\D^4 y
\frac{\de_\kappa\beta_\rho(x)}{(2\pi)^6}
\int\frac{\D^3\p}{2\varepsilon_p}\int\frac{\D^3\p'}{2\varepsilon_{p'}}
\delta^4\left(k-\frac{p+p'}{2} \right)\\
&\times \mathcal{T}^{\rho\kappa}(p,p')_{ab}
\E^{\I(p-p')(x-y)}\E^{z(p-p')\beta}n\ped{F}(p)(1-n\ped{F}(p')),
\end{split}
\end{equation*}
and
\begin{equation*}
\begin{split}
\Delta_B W_{ab}^+(x,k)=&-\int_0^1\D z\int_{\Sigma\ped{B}}\D\Sigma_\lambda(y)
\de_\kappa\beta_\rho(x)\frac{(y-x)^\kappa}{(2\pi)^6}
\int\frac{\D^3\p}{2\varepsilon_p}\int\frac{\D^3\p'}{2\varepsilon_{p'}}
\delta^4\left(k-\frac{p+p'}{2} \right)\\
&\times \mathcal{T}^{\lambda\rho}(p,p')_{ab}
\E^{\I(p-p')(x-y)}\E^{z(p-p')\beta}n\ped{F}(p)(1-n\ped{F}(p')).
\end{split}
\end{equation*}
Consider the 4D integral first. Assuming that the region $\Omega_B$ is large enough, we can
approximate it with:
\begin{equation*}
\begin{split}
\int_{\Omega_B}\D^4 y\, \E^{\I(p-p')(x-y)}\simeq\delta t
(2\pi)^3 \delta^3\left(\bf{p}-\bf{p}' \right),
\end{split}
\end{equation*}
where $\delta t$ is the temporal extent of the region $\Omega_B$. Hence:
\begin{equation}
\label{eq:DeltaVWig}
\begin{split}
\Delta_\Omega W_{ab}^+(x,k)=&-\delta t\,
\de_\kappa\beta_\rho(x)
\frac{1}{(2\pi)^3}\int\frac{\D^3\p}{4\varepsilon_p^2} \delta^4\left(k-p \right)
\mathcal{T}^{\rho\kappa}(p,p)_{ab} n\ped{F}(p)(1-n\ped{F}(p)).
\end{split}
\end{equation}
Plugging this expression in the \eqref{eq:PolVandB}, taking into account that 
$$
\tr\left[\gamma^\mu\gamma^5(\slashed{p}'+m)\gamma^\lambda(\slashed{p}+m)\right]
=4\I\epsilon^{\mu\lambda\tau\sigma}p_\tau p'_\sigma,
$$
and keeping in mind the \eqref{tb}, we readily find that
$\tr\left[\gamma^\mu\gamma^5 \Delta_\Omega W^+(x,k)\right]=0$,
so that the 4D integral does not contribute to the spin vector. We note that such
a result is naturally expected for the angular momentum-boost operators, as they are a conserved 
charge independent of the integration hypersurface, but it is not obvious for the $\wQ_x$ 
pseudo-tensor. Indeed, the vanishing of the volume term is a specific result owing 
to the choice of the region where the Gauss theorem has been applied and the observable, the spin 
polarization vector.

We can now move on to the $\Sigma_B$ term. We start to evaluate the numerator:
\begin{equation*}
\begin{split}
\mathcal{N}^\mu \equiv & \int_\Sigma \di\Sigma \cdot k\,\tr\left[\gamma^\mu\gamma^5 \Delta_B W^+(x,k)\right]\\
=&\frac{-2}{(2\pi)^6}\int_\Sigma \D\Sigma\cdot  k
\int_0^1\D z \int\frac{\D^3\p}{2\varepsilon_p}\int\frac{\D^3\p'}{2\varepsilon_{p'}}
\delta^4\left(k-\frac{p+p'}{2} \right)
\int_{\Sigma\ped{B}}\D\Sigma_\lambda(y)\de_\kappa\beta_\rho(x)(y-x)^\kappa \E^{\I(p-p')(x-y)}\\
&\times \left[\I\epsilon^{\mu\lambda\tau\sigma}p_\tau p'_\sigma k^\rho +
\I\epsilon^{\mu\rho\tau\sigma}p_\tau p'_\sigma k^\lambda \right]
\E^{z(p-p')\beta}n\ped{F}(p)(1-n\ped{F}(p'))
\end{split}
\end{equation*}
and the denominator:
\begin{equation*}
\mathcal{D} \equiv \int_\Sigma \D\Sigma \cdot  k\, \tr\left[W^+_0(x,k)\right]
    =\frac{4m}{(2\pi)^3}\int\D\Sigma\cdot k\, \delta(k^2-m^2)\theta(k_0)n\ped{F}(k)
\end{equation*}
of the second term on the right hand side of \eqref{eq:PolVandB}. If the hypersurface $\Sigma_B$ is 
large compared to the other scales, it can be approximated with an unbounded hyperplane and 
so:
\begin{equation}\label{eq:Baseint}
\begin{split}
\int_{\Sigma\ped{B}}\D\Sigma_\lambda(y) (y-x)^\kappa \E^{\I(p-p')(x-y)} = &
    \int_{\Sigma\ped{B}}\D^3 y\,\hat{t}_\lambda (y-x)^\kappa \E^{\I(p-p')(x-y)}\\
\simeq& -\I\hat{t}_\lambda\Delta^\kappa_{\;\kappa'}(2\pi)^3\frac{\de}{\de p'_{\kappa'}}\delta^3(\vec{p}-\vec{p}')
    +\hat{t}_\lambda\hat{t}^\kappa (2\pi)^3 \Delta t \delta^3(\vec{p}-\vec{p}'),
\end{split}
\end{equation}
where $\hat{t}$ is the unit vector normal to $\Sigma\ped{B}$ (which corresponds to the 
time direction in the QGP frame of figure~\ref{figure}) and 
$\Delta^{\mu\nu}=\eta^{\mu\nu}-\hat{t}^\mu \hat{t}^\nu$, $\Delta t = (y-x)\cdot \hat{t}$
and $y\cdot\hat{t}$ is constant in $\Sigma\ped{B}$ by definition. Notice that both terms in the 
integral of $\mathcal{N}$ contain the momenta $p$ and $p'$ contracted with the Levi-Civita tensor 
and that the second term in~\eqref{eq:Baseint} sets $p'=p$ in $\mathcal{N}$ after integrating $p'$; 
we therefore conclude that the second term in~\eqref{eq:Baseint} does not bring any contribution.
Therefore:
\begin{equation*}
\begin{split}
\mathcal{N}^\mu \simeq & -\frac{2}{(2\pi)^3}\int_\Sigma \D\Sigma\cdot  k \int_0^1\D z 
\int\frac{\D^3\p}{2\varepsilon_p}\int\frac{\D^3\p'}{2\varepsilon_{p'}} \delta^4\left(k-\frac{p+p'}{2} \right)
\de_\kappa\beta_\rho(x)\hat{t}_\lambda\Delta^\kappa_{\;\kappa'}
\frac{\de}{\de p'_{\kappa'}}\delta^3(\vec{p}-\vec{p}')\\
&\times \left[\epsilon^{\mu\lambda\tau\sigma}p_\tau p'_\sigma k^\rho +
\epsilon^{\mu\rho\tau\sigma}p_\tau p'_\sigma k^\lambda \right] \e^{z(p-p')\beta}n\ped{F}(p)(1-n\ped{F}(p')).
\end{split}
\end{equation*}
Integrating by parts in $p^\prime$ and taking advantage of the vanishing of the square bracket 
for $p'=p$, we obtain:
\begin{equation*}
\begin{split}
\mathcal{N}^\mu
=& \frac{1}{(2\pi)^3}\int_\Sigma \D\Sigma\cdot  k\,\de_\kappa\beta_\rho(x) 
    \int\frac{\D^3\p}{2\varepsilon_p^2}\delta^4(k-p)n\ped{F}(p)(1-n\ped{F}(p))
    \hat{t}_\lambda\Delta^\kappa_{\;\kappa'}
    \left[\epsilon^{\mu\lambda\tau\sigma}p_\tau \frac{\de p_\sigma}{\de p_{\kappa'}} k^\rho +
    \epsilon^{\mu\rho\tau\sigma}p_\tau \frac{\de p_\sigma}{\de p_{\kappa'}} k^\lambda \right],
\end{split}
\end{equation*}
where we have to take into account that $\partial p_0/\partial p_{\kappa'}$ is non-trivial 
being $p$ on-shell. We can then integrate in $p$ getting:
\begin{equation*}
\begin{split}
\mathcal{N}^\mu=& \frac{1}{(2\pi)^3}\int_\Sigma \D\Sigma\cdot  k\,\de_\kappa\beta_\rho(x) 
    \theta(k_0)\delta(k^2-m^2)n\ped{F}(k)(1-n\ped{F}(k)) \,
    \hat{t}_\lambda\Delta^\kappa_{\;\kappa'} \frac{\de k_\sigma}{\de k_{\kappa'}}
    \frac{k_\tau}{\varepsilon_k}\left[\epsilon^{\mu\lambda\tau\sigma}  k^\rho +
    \epsilon^{\mu\rho\tau\sigma} k^\lambda \right],
\end{split}
\end{equation*}
where $k$ is on-shell.
Notice that the index $\kappa'$ is in fact a spatial index, therefore:
\begin{equation*}
\frac{\de k_\sigma}{\de k_{\kappa'}}=\Delta_\sigma^{\,\,\kappa'}
    -\hat{t}_\sigma\frac{k^{\kappa'}}{\varepsilon_k}.
\end{equation*}
Using the previous derivative and the decompositions
\begin{equation*}
k^\rho=\hat{t}^\rho \varepsilon_k + \Delta^\rho_{\,\,\rho'}k^{\rho'},\quad
\epsilon^{\mu\rho\tau\sigma}=\hat{t}^\rho \epsilon^{\mu\rho'\tau\sigma}\hat{t}_{\rho'}
    + \epsilon^{\mu\rho'\tau\sigma}\Delta^\rho_{\,\,\rho'}\, ,
\end{equation*}
we can rewrite $\mathcal{N}$ as:
\begin{equation*}
\begin{split}
\mathcal{N}^\mu=& \frac{1}{(2\pi)^3}\int_\Sigma \D\Sigma\cdot  k\,\de_\kappa\beta_\rho(x) 
    \theta(k_0)\delta(k^2-m^2)n\ped{F}(k)(1-n\ped{F}(k))\\
    &\times\epsilon^{\mu\lambda\tau\sigma}\frac{k_\tau}{\varepsilon_k}\left(
    2\,\hat{t}^\rho \hat{t}_\lambda \Delta^\kappa_{\,\,\sigma} \varepsilon_k
    +\Delta^\rho_{\,\,\kappa'}k^{\kappa'} \Delta^\kappa_{\,\,\sigma} \hat{t}_\lambda
    +\Delta^\rho_{\,\,\sigma} \Delta^\kappa_{\,\,\kappa'}k^{\kappa'}\hat{t}_\lambda 
    +\Delta^\rho_{\,\,\lambda}\Delta^\kappa_{\,\,\sigma}\varepsilon_k\right).
\end{split}
\end{equation*}

Now we can split the gradient of $\beta$ as the sum of the thermal vorticity and the 
thermal shear:
\begin{equation*}
\de_\kappa\beta_\rho = \frac{1}{2}\left[\de_\kappa\beta_\rho+\de_\rho\beta_\kappa\right]
    -\frac{1}{2}\left[\de_\rho\beta_\kappa-\de_\kappa\beta_\rho\right]
    =\xi_{\kappa\rho}-\varpi_{\kappa\rho}.
\end{equation*}
thus obtaining the linear contributions from the angular momentum-boost operators $\wJ_x$
and the operator $\wQ_x$ in the~\eqref{rholeappr}. For the thermal vorticity $\varpi$ we obtain:
\begin{equation*}
\begin{split}
\mathcal{N}^\mu_\varpi=& \frac{-1}{(2\pi)^3}\int_\Sigma \D\Sigma\cdot  k\,
    \theta(k_0)\delta(k^2-m^2)n\ped{F}(k)(1-n\ped{F}(k))
    \epsilon^{\mu\nu\sigma\tau} \varpi_{\nu\sigma} k_\tau
\end{split}
\end{equation*}
whence:
\begin{equation*}
 S_\varpi^\mu(k)= -\frac{1}{8m} \epsilon^{\mu\nu\sigma\tau} k_\tau 
 \frac{\int_{\Sigma} \di \Sigma \cdot k \, n_F (1 -n_F) 
 \varpi_{\nu\sigma}}{\int_{\Sigma} \di \Sigma \cdot k \, n_F},
\end{equation*}
which is the known expression in eq.~\eqref{basic}. Similarly for the thermal shear term, simple 
calculations yield:
\begin{equation*}
\begin{split}
\mathcal{N}^\mu_\xi=& -\frac{1}{(2\pi)^3}\int_\Sigma \D\Sigma\cdot  k\,
    \theta(k_0)\delta(k^2-m^2)n\ped{F}(k)(1-n\ped{F}(k))
    \epsilon^{\mu\nu\sigma\tau} k_\tau \hat{t}_\nu \xi_{\sigma\rho} \frac{k^\rho}{\varepsilon_k}
\end{split}
\end{equation*}
whence it follows, replacing the four-momentum $k$ with $p$:
\begin{equation}\label{main}
\boxed{
S_\xi^\mu(p)=-\frac{1}{4m} \epsilon^{\mu\nu\sigma\tau} \frac{p_\tau p^\rho}{\varepsilon}
 \frac{\int_{\Sigma} \di \Sigma \cdot p \; n_F (1 -n_F) 
 \hat{t}_\nu\xi_{\sigma\rho}}{\int_{\Sigma} \di \Sigma \cdot p \; n_F},}
\end{equation}
which is, in general, non-vanishing. As has been mentioned, the integration hypersurface $\Sigma$, 
in relativistic nuclear collisions, is the freeze-out $\Sigma_{FO}$ (see figure~\ref{figure}). 
This term is a new, non-dissipative contribution to the spin polarization vector at local thermodynamic 
equilibrium to be added to the \eqref{basic}. 

\section{Discussion and conclusions}

The striking difference between the \eqref{basic} and the new term \eqref{main} is that 
the latter apparently breaks covariance, for the presence of the $\hat t_\nu$, the time direction
in the QGP frame (see figure~\ref{figure}). The reason is the inevitable dependence of $\wQ_x$ 
operator \eqref{quadr} on the particular 3D hypersurface of integration, as it has been discussed
in section~\ref{sec:local}. The appearance of a particular vector $\hat t$, which is, in a sense, 
the best approximation of the unit vector perpendicular to the hypersurface $\Sigma_{FO}$, is the 
telltale sign of this dependence.

The tensor $\xi$ can be covariantly decomposed along the four-velocity of the fluid into: 
\be\label{xidecomp}
 \xi_{\sigma\rho} = \frac{1}{2} \partial_\sigma \left( \frac{1}{T} \right) u_\rho +
 \frac{1}{2} \partial_\rho \left( \frac{1}{T} \right) u_\sigma  + \frac{1}{2T} \left( A_\rho u_\sigma +
   A_\sigma u_\rho \right) + \frac{1}{T} \sigma_{\rho\sigma} + \frac{1}{3T} \theta \Delta_{\rho\sigma}\, .
\ee
In the \eqref{xidecomp}, denoting by $\nabla_\mu = \partial_\mu - u_\mu u \cdot \partial$ and
$\Delta_{\mu\nu} = g_{\mu\nu} - u_\mu u_\nu$, $A = u \cdot \partial u$ is the acceleration field, 
$\sigma$ is the properly called shear tensor:
$$
  \sigma_{\mu\nu} = \frac{1}{2} (\nabla_\mu u_\nu + \nabla_\nu u_\mu) - \frac{1}{3} \Delta_{\mu\nu} \theta
$$
and $\theta = \nabla \cdot u$ the expansion scalar. All of these terms can contribute to the spin vector 
\eqref{main}, however only one of them, namely the temperature gradient, has a non-vanishing non-relativistic 
limit:
\be\label{nrlimit}
  {\bf S}_\xi = \frac{1}{8} {\bf v} \times 
   \frac{\int \di^3 {\rm x} \; n_F (1-n_F) \nabla  \left( \frac{1}{T} \right)}{\int \di^3 {\rm x} \; n_F } 
\ee
while all remaining terms, including the spin-shear coupling, are purely relativistic. Such a
term was already noticed in the non-relativistic limit of thermal vorticity \cite{Becattini:2017vsh}, 
and provides an equal contribution. This will be the subject of further work.

The additional term of spin polarization vector \eqref{main} is linear in the gradients of the thermodynamic
fields and can then play a major role in driving the local spin polarization pattern in 
relativistic heavy ion collisions. We will show its numerical impact in a forthcoming paper 
\cite{Becattini:2021iol}. 

\acknowledgments
While we were finalizing this work, we got to know of a simultaneous similar study \cite{Liu:2021uhn}. 
M.B. is supported by the Florence University fellowship {\em Effetti quantistici nei fluidi
relativistici}.
\appendix

\section{Canonical stress-energy tensor and operator  \texorpdfstring{$\wQ_x$}{Qx}}
\label{sec:canonical}

The purpose of this section is to show that, for the free Dirac field, the operator $\wQ_x$ in 
eq.~\eqref{quadr} with the Belinfante stress-energy tensor:
\be\label{quadbel}
\wQ^{\lambda\nu}_x = \int_\Sigma \di \Sigma_\mu \; \left[(y-x)^\lambda \wT_B^{\mu\nu}(y) + 
  (y-x)^\nu \wT_B^{\mu\lambda}(y)\right]
\ee
is the same if we replaced $\wT_B$ with the canonical stress-energy tensor $\wT_C$ of the 
free Dirac field. Starting from the pseudo-gauge transformation relation between $\wT_B$ and $\wT_C$:
$$
 \wT_B^{\mu\nu} = \wT_C^{\mu\nu} + \frac{1}{2} \partial_\alpha \left( \wspt^{\alpha\mu\nu}
- \wspt^{\mu\alpha\nu} - \wspt^{\nu\alpha\mu} \right),
$$
where $\wspt$ is the {\em canonical} spin tensor of the free Dirac field. This is known to be 
dual to the axial current, hence it is completely antisymmetric in the three indices; hence, the
above transformation formula simplifies to:
$$
 \wT_B^{\mu\nu} = \wT_C^{\mu\nu} + \frac{1}{2} \partial_\alpha \wspt^{\alpha\mu\nu} .
$$
By plugging this formula into $\wQ_x$, we obtain:
\be\label{quadcan}
\wQ^{\lambda\nu}_x = \int_\Sigma \di \Sigma_\mu \; \left[(y-x)^\lambda \wT_C^{\mu\nu}(y) + 
  (y-x)^\nu \wT_C^{\mu\lambda}(y)\right] + \frac{1}{2}
  \int_\Sigma \di \Sigma_\mu \; \left[(y-x)^\lambda \partial_\alpha \wspt^{\alpha\mu\nu}
  + (y-x)^\nu \partial_\alpha \wspt^{\alpha\mu\lambda} \right] .
\ee
We now focus on the integral term involving the spin tensor. Integrating by parts we get:
$$
 \int_\Sigma \di \Sigma_\mu \; \left[(y-x)^\lambda \partial_\alpha \wspt^{\alpha\mu\nu}
  + (y-x)^\nu \partial_\alpha \wspt^{\alpha\mu\lambda} \right] =
  \int_\Sigma \di \Sigma_\mu \; \partial_\alpha \left[(y-x)^\lambda \wspt^{\alpha\mu\nu}
  + (y-x)^\nu \wspt^{\alpha\mu\lambda} \right] -   
  \int_\Sigma \di \Sigma_\mu \; \left( \wspt^{\lambda\mu\nu} + \wspt^{\nu\mu\lambda} \right) .
$$
The second integral on the right hand side vanishes because of the anti-symmetry of indices, 
so we are left with a 3D integral which can be turned into a surface 2D integral by means of 
the relativistic Stokes theorem:
$$
  \int_\Sigma \di \Sigma_\mu \; \partial_\alpha \left[(y-x)^\lambda \wspt^{\alpha\mu\nu}
  + (y-x)^\nu \wspt^{\alpha\mu\lambda} \right] = 
  \int_{\partial \Sigma} \di \tilde S_{\mu\alpha} \; \left[ (y-x)^\lambda \wspt^{\alpha\mu\nu}
  + (y-x)^\nu \wspt^{\alpha\mu\lambda} \right] .
$$
This integral can be made vanishing by enforcing suitable boundary conditions on
the Dirac field; if $\Sigma = \Sigma_{FO} \cup \sigma_\pm$ (see figure~\ref{figure}) 
and discussion in section~\ref{sec:local}), usually anti-periodic boundary conditions for
a compact hypersurface $\Sigma$ are enforced before taking the limit to infinity.  
Therefore, only the first integral in eq.~\eqref{quadcan} contributes to the 
operator $\wQ_x$, which is the same in form as in eq.~\eqref{quadbel}.

\section{Local thermal equilibrium value of Wigner function}
\label{sec:AppWigFun}

In this Appendix we evaluate the local thermal equilibrium Wigner function in eq.~\eqref{eq:WigFunLE},
which is given by the sum of the Wigner function at global homogeneous thermal equilibrium~\cite{hakim}
\begin{equation*}
W^+_0(x,k)=\mean{\h{W}^+(x,k)}_{\beta(x)}=\left(m+\gamma^\mu k_\mu\right)
    \delta(k^2-m^2)\theta(k_0)\frac{1}{(2\pi)^3} n\ped{F}\left( k\right)
\end{equation*}
and the first correction in temperature gradients~\eqref{eq:WigfunCorr}. First, to work out
the~\eqref{eq:WigfunCorr}, we take advantage of the following relation between
Belinfante stress-energy tensor operator and the Wigner operator:
\begin{equation*}
\h{T}^{\lambda\rho}(y)=\frac{1}{2}\int\D^4 k' \left(k^{\prime\rho}
    \tr\left[\gamma^\lambda \h{W}(y,k')\right]
    +k^{\prime\lambda} \tr\left[\gamma^\rho \h{W}(y,k')\right]\right),
\end{equation*}
hence the~\eqref{eq:WigfunCorr} becomes
\begin{equation}
\label{eq:DeltaWigApp}
\Delta W^+_{ab}(x,k) = -\frac{1}{2}\int_0^1\D z\int_{\Sigma}\D\Sigma_\lambda(y)
    \Delta\beta_\rho(x,y)\int\D^4 k' \sum_{cd}
    \left(k^{\prime\rho}\gamma^\lambda_{dc}+ k^{\prime\lambda}\gamma^\rho_{dc} \right)
    \mean{\h{W}^+_{ab}(x,k)\h{W}_{cd}(y+\I z\beta(x),k')}_{c,\beta(x)}.
\end{equation}

Thereby, the calculation of~\eqref{eq:WigfunCorr} boils down to the evaluation of the
correlator of two Wigner operators at homogeneous global equilibrium. For this purpose, we 
shall express the Wigner function in terms of the normal modes of the Dirac field:
\begin{equation*}
\Psi (x)=\sum_{\sigma=-1/2}^{1/2} \frac{1}{(2\pi)^{3/2}}\int\frac{\di^3 k}{2\varepsilon_k}
    \left[  u_\sigma(k)\e^{-\ii k\cdot x} \h{a}_\sigma(k)
    + v_\sigma(k)\e^{\ii k\cdot x} \h{b}^\dagger_\sigma(k)\right],
\end{equation*}
where creation and annihilation operators are covariantly normalized:
$$
 [\h{a}_{\sigma}(q),\h{a}_{\sigma'}^\dagger(q')]_\pm= 2 \varepsilon_q \,
 \delta_{\sigma\sigma'}\delta^3(\vec{q}-\vec{q}'),
$$
and $u$ and $v$ are the spinors of the Dirac field satisfying:
$$
 \bar u_\sigma(k) u_{\sigma'}(k)= 2m\delta_{\sigma\sigma'}, \qquad
 \bar v_\sigma(k) v_{\sigma'}(k)= -2m\delta_{\sigma\sigma'}.
$$
From the definition of Wigner operator~\eqref{eq:wigop} we obtain:
\begin{equation*}
\h{W}^+_{ab}(x,k)=\frac{1}{(2\pi)^3}\sum_{\tau,\tau'}\int\frac{\di^3 p}{2\varepsilon_p}
    \int\frac{\di^3 p'}{2\varepsilon_{p'}}\delta^4\left(k-\tfrac{p+p'}{2}\right)
    \e^{-\I x(p'-p)} u_{\tau'}(p')_a \bar{u}_{\tau}(p)_b
    \h{a}^\dagger_\tau(p) \h{a}_{\tau'}(p').
\end{equation*}
The thermal correlator between two Wigner operators involves the connected part of the thermal average
of four creation and annihilation operators which is given by:
\begin{equation*}
\mean{\h{a}^\dagger_1 \h{a}_2 \h{a}^\dagger_3 \h{a}_4}_c=
    \mean{\h{a}^\dagger_1 \h{a}_2 \h{a}^\dagger_3 \h{a}_4}
        -\mean{\h{a}^\dagger_1 \h{a}_2}\mean{ \h{a}^\dagger_3 \h{a}_4}
    =\mean{\h{a}^\dagger_1 \h{a}_4}\mean{ \h{a}_2\h{a}^\dagger_3 },
\end{equation*}
where $\h{a}^\dagger_1$ and $\h{a}_2$ come from the first Wigner operator and $\h{a}^\dagger_3$ and $\h{a}_4$ from the
second one. The thermal averages of creation-annihilation operators at the homogeneous 
thermodynamic equilibrium are well known quantities:
\begin{equation*}
\begin{split}
\mean{\h{a}^\dagger_\tau(k) \h{a}_{\sigma}(q)}_{\beta(x)}=&
    \delta_{\tau\sigma}2\varepsilon_q \delta^3(\vec{k}-\vec{q})n\ped{F}(k),\\
\mean{\h{a}_{\tau'}(k')\h{a}^\dagger_{\sigma'}(q')}_{\beta(x)}=&
    \delta_{\tau'\sigma'}2\varepsilon_{q'}\delta^3(\vec{k'}-\vec{q'})(1-n\ped{F}(k')),
\end{split}
\end{equation*}
while all other combinations vanish. Taking advantage of the spinor identity
\begin{equation*}
\sum_\sigma u_\sigma(p) \bar{u}_\sigma(p)=\slashed{p}+m,
\end{equation*}
it is then straightforward to show that the thermal correlation between two Wigner operators 
turns out to be:
\begin{equation*}
\begin{split}
\mean{\h{W}^+_{ab}(x,k)\h{W}_{cd}(y+\I z\beta(x),k')}_{c,\beta(x)}= &
    \frac{1}{(2\pi)^6}\int\frac{\D^3\p}{2\varepsilon_p}\int\frac{\D^3\p'}{2\varepsilon_{p'}}
    \delta^4\left(k-\frac{p+p'}{2} \right)\delta^4\left(k'-\frac{p+p'}{2} \right)\times\\
    & (\slashed{p}'+m)_{ad}(\slashed{p}+m)_{cb}
    \E^{\I(p-p')(x-y)}\E^{z(p-p')\beta}n\ped{F}(p)(1-n\ped{F}(p')).
\end{split}
\end{equation*}
Plugging this expression into~\eqref{eq:DeltaWigApp} we readily obtain the eq.~\eqref{eq:DeltaWigPriv}.

\bibliographystyle{apsrev4-1}

\end{document}